# Emergence of non-Fermi liquid behaviors in 5*d* perovskite SrIrO$_3$ thin films: interplay between correlation, disorder, and spin-orbit coupling

Abhijit Biswas[a], Ki-Seok Kim[a,b], and Yoon H. Jeong[a,*]

[a]*Department of Physics, POSTECH, Pohang, 790-784, South Korea*
[b]*Institute of Edge of Theoretical Science (IES), POSTECH, Pohang, 790-784, South Korea*
[*] Corresponding author. *E-mail address:* yhj@postech.ac.kr

**Abstract**

We investigate the effects of compressive strain on the electrical resistivity of 5*d* iridium based perovskite SrIrO$_3$ by depositing epitaxial films of thickness 35 nm on various substrates such as GdScO$_3$ (110), DyScO$_3$ (110), and SrTiO$_3$ (001). Surprisingly, we find anomalous transport behaviors as expressed by $\rho \propto T^\varepsilon$ in the temperature dependent resistivity, where the temperature exponent $\varepsilon$ evolves continuously from 4/5 to 1 and to 3/2 with an increase of compressive strain. Furthermore, magnetoresistance always remains positive irrespective of resistivity upturns at low temperatures. These observations imply that the delicate interplay between correlation and disorder in the presence of strong spin-orbit coupling is responsible for the emergence of the non-Fermi liquid behaviors in 5*d* perovskite SrIrO$_3$ thin films. We offer a theoretical framework for the interpretation of the experimental results.

*Keywords:* SrIrO$_3$ thin films, Strain effect, Disorder, Metal-insulator transition, Non-Fermi liquid behavior

# Highlights

- We studied the effect of compressive strain on the perovskite SrIrO$_3$ thin films.
- We revealed non-Fermi liquid behaviors in the transport properties.
- Irrespective of weak localization effects, magnetoresistance remains positive.
- We interpret the anomalous transport properties as arising from the interplay between correlation, disorder, and spin-orbit coupling.

## 1. Introduction

Understainding the microscopic origin of non-Fermi liquid physics has been one of the long-standing problems in strongly correlated systems [1]. Although various mechanisms were provided to explain anomalous non-Fermi liquid transport phenomena, it is believed that for many strongly correlated systems the non-Fermi liquid behaviors are most likely triggered by the interplay of quenched disorder and strong electronic correlation [2]. This non-Fermi liquid phenomenology is rather ubiquitous near metal-insulator transitions [3]. Here, we wish to report non-Fermi liquid (NFL) behaviors over a wide temperature range near the metal-insulator transition (MIT) found in 5$d$ element perovskite (Pv) SrIrO$_3$ thin films. We interpret that these NFL behaviors are the emerging result from a delicate interplay between correlation and disorder in the presence of strong spin-orbit coupling (SOC). Pv-SrIrO$_3$ would be an example of strong spin-orbit coupled correlated systems.

5$d$ transition metal oxides (TMO) have become a topic of intense activities in condensed matter physics following the discovery of $J_{\text{eff}}$ = 1/2 Mott insulating ground state in Sr$_2$IrO$_4$

**[4,5]**. In fact, many intriguing properties such as correlated insulator, charge-density wave, Weyl semimetal and possible topological insulator have been studied in 5$d$ TMOs by taking into account the energy scale of large SOC **[6]**. Among 5$d$ oxides, probably Ruddlesden-Popper series $Sr_{n+1}Ir_nO_{3n+1}$ ($n$ = 1, 2, and ∞) have been the most investigated materials as they show a rich phase diagram involving MITs with increasing $n$ **[7]**. With increasing $n$, i.e. increasing the number of $IrO_2$ planes, the bandwidth of the Ir 5$d$ band becomes broader, and as a result Pv-$SrIrO_3$ ($n$ = ∞) becomes a correlated semimetal **[8-11]**. This novel phenomenon is due to the interplay between local Coulomb interaction ($U$) and strong SOC as the strength of SOC can be as large as 0.3 ~ 0.5 eV in 5$d$ oxides (SOC ∝ $z^4$ where $z$ is the atomic number), comparable to the corresponding bandwidth or Coulomb interaction, and thus play a decisive role in the physics of 5$d$ oxides **[6]**.

From the viewpoint of crystal structures, the ground state of $SrIrO_3$ at room temperature and atmospheric pressure is the structure of hexagonal $BaTiO_3$ **[12]**. Studies on hexagonal $SrIrO_3$ single crystals revealed that it is a metallic system exhibiting NFL behaviors; the specific heat displays a typical logarithmic divergence and the electrical resistivity follows $T^{3/2}$ **[13]**. This NFL physics has been attributed to quantum criticality. Pv-$SrIrO_3$ can be obtained only at elevated pressure (40 kbar) and temperature (1000 °C) and is a paramagnetic metal **[12,14,15]**. In fact, Pv-$SrIrO_3$ is supposed to be a correlated bad metal, i.e., mean free path comparable to the inter-atomic distance, and is presumably close to a metal-insulator phase boundary **[7]**. Thus, if Pv-$SrIrO_3$ can be stabilized at room temperature, it would open up a possibility of investigating new emergent phenomena because Pv-$SrIrO_3$ would be susceptible to external perturbations due to its vicinity to MIT. In order to obtain Pv-$SrIrO_3$ and also to induce new physical phenomena different from bulk properties, we resorted to thin film technology, where underlying substrates help stabilize the perovskite phase. We demonstrate that Pv-SrIrO$_3$ thin films display NFL behaviors and, more surprisingly, with an increase of compressive strain, the temperature dependence of the resistivity as expressed $\rho \propto T^\varepsilon$ evolves from that with $\varepsilon = 4/5$ to 1 and to 3/2, due to a subtle interplay between correlation, disorder, and SOC.

## 2. Materials and methods

While bulk Pv-SrIrO$_3$ is metastable at room temperature, the metastable phase can be stabilized with thin film growth, and, in addition, key parameters of Pv-SrIrO$_3$ such as band width and correlation can be tuned with strained thin films by changing underlying substrates. We have grown Pv-SrIrO$_3$ thin films (typical thickness ~ 35 nm) on various lattice mismatched substrates such as GdScO$_3$ (110), DyScO$_3$ (110), and SrTiO$_3$ (001) **[9]**. We used pulsed laser deposition (KrF laser with $\lambda$ = 248 nm) to grow SrIrO$_3$ thin films from a polycrystalline target. The target was prepared by a solid-state reaction method; stoichiometric mixing of SrCO$_3$ and IrO$_2$ raw powders is followed by sintering at 1000º C for 48 hrs. The laser was operated at frequency 4 Hz, and the substrate temperature and oxygen partial pressure were 550 ºC and 20 mTorr, respectively. The target to substrate distance was kept ~50 mm. After growth, all the films were annealed at the same oxygen partial pressure and temperature to compensate for any oxygen deficiency. Crystalline quality of the films was checked by X-ray diffraction (XRD) measurements; XRD was carried out with the Empyrean XRD System from PANalytical. Four-probe van der Pauw geometry was used for electrical transport measurements. For magnetoresistance measurements, the applied magnetic field strength was -9 T $\leq B \leq$ 9 T in the out-of-plane direction.

## 3. Results and discussion

Structurally, bulk Pv-SrIrO$_3$ is of orthorhombic structure ($a$ = 5.60 Å, $b$ = 5.58 Å, $c$ = 7.89 Å) **[12]**. Note that GdScO$_3$ (110) and DyScO$_3$ (110) substrates are also orthorhombic and orthorhombic indices are typically used to indicate their orientation. SrTiO$_3$ (001) substrates, on the other hand, are cubic. For thin films on substrates, pseudo-cubic indices are preferred for easy comparison. If the *pseudo-cubic* ($a_{pc}$) lattice parameter is converted from the orthorhombic lattice parameters of bulk Pv-SrIrO$_3$, we obtain $a_{pc}$ ~ 3.96 Å. This would match very well GdScO$_3$ substrates with $a_{pc}$ ~ 3.96 Å, correspond to +0.50% compressive strain for the films on DyScO$_3$ substrates with $a_{pc}$ ~ 3.94 Å, and +1.54% compression for those on SrTiO$_3$ substrates with $a_{pc}$ ~ 3.90 Å as illustrated in Fig. 1a. The epitaxial nature of the perovskite thin films were confirmed by XRD measurements. The XRD analysis ($\theta$-$2\theta$ scan) of the Pv-SrIrO$_3$ films grown on aforementioned substrates show crystalline peaks without any impurity or additional peaks; for clarity, only the low angle data including the $(001)_{pc}$ peak are shown in Fig. 1b. The films on all the substrates exhibit clear layer thickness fringes (results of coherent scattering from a finite number of lattice planes with thickness of the film), showing the surface smoothness as well as high crystalline quality. AFM characterization of the film surfaces also confirmed that they were nearly flat with maximum roughness of 1.5 nm (not shown here). From Fig. 1b, it is noted that with an increasing lattice mismatch, the $2\theta$ value of the film peak maximum decreases and thus the film's out-of-plane lattice constant increases correspondingly. This fact indicates the in-plane film lattice constant is locked with that of the underlying substrate and indeed the films are under compression **[9]**.

Having confirmed the epitaxial quality, we then measured resistivity ($\rho$) of the Pv-

SrIrO$_3$ films grown on various lattice-mismatched substrates as shown in Fig. 2. The film on the best lattice-matched substrate GdScO$_3$ shows $\rho \sim 1.45$ mΩ·cm at $T$ = 300 K, which is much smaller than that of the bulk polycrystalline Pv-SrIrO$_3$ $\rho \sim 4$ mΩ·cm at $T$ = 300 K **[12]**. Obviously, the resistivity of bulk polycrystalline Pv-SrIrO$_3$ was affected by numerous grain boundaries and large porosity. The metallicity of the film on GdScO$_3$ persists down to the lowest temperature $T$ = 2 K and the change in the resistivity ratio $\rho/\rho_{300K}$ was small $\sim 0.7$ as seen in Fig. 2. This signifies a semi-metal-like characteristic of the Pv-SrIrO$_3$ film. Hall effect measurements also confirmed the semimetallic state as the carrier concentration was found to be ~10$^{20}$ cm$^{-3}$ at $T$ = 300 K **[9]**. This semimetallic nature indicates that the band gap is not open and the Fermi level may be located inside a pseudo-gap of the correlation split lower Hubbard band and upper Hubbard band of the $J_{eff}$ = 1/2 band, where only one electron would occupy for Ir$^{4+}$ ($d^5$). Imposing *compressive* strain by growing films on DyScO$_3$ as well as on SrTiO$_3$ brought about an increase in resistivity with respect to the best lattice-matched film grown on GdScO$_3$. The values of the room temperature resistivity are 1.45 mΩ·cm, 1.82 mΩ·cm, and 2.05 mΩ·cm for the films on GdScO$_3$, DyScO$_3$, and SrTiO$_3$, respectively. This resistivity variation under compression can be understood in terms of the bandwidth reduction brought by compression. As the compressive strain in the film would change both *Ir-O-Ir* bond angle and *Ir-O* bond length, the associated change in the electronic bandwidth (*W*) is given by $W \propto \frac{\cos\varphi}{d^{3.5}}$, where $d$ is the *Ir-O* bond length and $\varphi = (\pi - \theta)/2$ is the buckling deviation of the *Ir-O-Ir* bond angle $\theta$ from π **[9]**. Thus, the buckling would cause a band narrowing and less mobility of the carriers in the band.

The most striking feature in the electrical resistivity for the compressively strained film is the temerpature dependence as expressed by $\rho \propto T^\varepsilon$. Fitting the temperature depenent resistivity data to the power law, we found that in the metallic region the transport property displays a NFL behavior with $\varepsilon = 4/5$ for the films on GdScO$_3$, $\varepsilon = 1$ for the films on DyScO$_3$, and $\varepsilon = 3/2$ for the films on SrTiO$_3$. Fig. 3(a)-(c) depict the situation clearly. Thus, with an increase of the compressive strain, the temerpature exponent $\varepsilon$ changes from 4/5 to 1 and to 3/2. It may be noted that the films on DyScO$_3$ and SrTiO$_3$ substrates also show resistivity upturns at $T = 20$ K and $T = 50$ K, respectively, and continue to increase with decreasing temperature. This feature indicates that disorder and weak localization affect the transport properties of the Pv-SrIrO$_3$ films. It turns out that these upturns go well with the two-dimensional localization theory due to disorder, that is, $\sigma \propto \ln T$ as shown in Fig. 4(a)-(b) **[16]**. This certainly appears surprising because two-dimensional localization emerges in the three-dimensional films. Table 1 summarizes the temperature dependent resistivity in the whole temperature range. Based on the evolution of the temperature exponent of the resistivity variations with an increase in compressive strain and the seemingly contradictory two-dimensional localization for the three-dimensional films, it can be said that a change in correlation as well as the presence of disorder should be taken into account for the transport phenomena. Indeed, we have shown in our recent study that increasing the compressive-strain further in Pv-SrIrO$_3$ films renders disorder more effective in the films and eventually induces a transition to a fully insulating state **[9]**. Recent resistivity measurements by several other groups **[8,10]** also reported the strain-dependent behaviors and metal-insulator transitions in Pv-SrIrO$_3$ films, emphasizing the role of disorder.

Possible interplay between disorder and correlation in the compressively strained Pv-SrIrO$_3$ films was further confimed by magnetoresistance (MR) measurements. MR is defined as $\frac{\rho(B)-\rho(0)}{\rho(0)}$, i.e., a change in electric resistivity under the influence of an external magnetic field *B*. Along with other intrinsic physical properties, MR is extremely useful in obtaining important clues to the underlying state: (1) for magnetically ordered films, below magnetic ordering temperatures MR is often found to be negative due to reduced magnetic scattering as the moments are ordered **[17]**, (2) a simple paramagnetic metal shows positive MR due to the classical Lorentz contribution to the orbital motion of electrons **[18]**, (3) in the case of a metallic sample where a resistivity upturn happens at low temperatures, MR would be negative due to weak localization related with disorder effects **[19]**. For the present compressively strained films, MR remains always positive at low temperatures, irrespective of any unusual features in resistivity, with quadratic field dependence as shown in Fig. 5. This surprising positive MR even in the presence of localization effects seems to imply that for the compressively strained films, the interplay of disorder and correlation, not either disorder or correlation alone, along with SOC have to play a role in the electrical transport. The positive MR also removes a possibility of any long-range magnetic ordering. The absence of magnetic ordering was further confirmed by the magnetization measurements (not shown here), which revealed paramagnetic behaviors in the whole temperature range. Even in the absence of long range ordering, however, we cannot exclude a possibility of the presence of local moments which might play a role in the temperature dependent resistivity variations. While Pv-SrIrO$_3$ is certainly a paramagnetic system, local magnetic moments or small magnetic clusters are still possible to influence the electronic transport significantly in the presence of disorder. At this point, it is of value to compare the transport properties of Pv-SrIrO$_3$ in two contrasting cases, that is, the films with varying thickness on a given substrate and those on various substrates with constant thickness. The former case was previously reported [9] and the latter is the present situation. When the thickness of the films on a given substrate (GdScO$_3$) is reduced, the resistivity shows characteristic behaviors, distinct from the present strain dependent case: (1) the power law exponent ($\rho \propto T^{4/5}$) remains the same regardless of thickness. (2) Weak localization effects are seen in thin films below 10 nm thickness at low temperatures. (3) MR at low temperatures are negative. (4) When the thickness is reduced from 4 nm to 3 nm, an MIT occurs and the resistivity of the insulating film is described by variable range hopping. Thus, from these features, we are able to conclude that effective disorder increases as thickness is reduced and the thickness dependent MIT is disorder driven [9].

In order to provide a theoretical understanding of the present experimental findings, it is pointed out that the continuous change of the temperature exponent $\varepsilon$ in electrical resistivity reminds us of the fundamental problem on the nature of the NFL physics near the MIT (as compressive strain eventually brings about MIT for Pv-SrIrO$_3$ films **[9]**). We conjecture that our observations imply a particular type of interplays between correlation of electrons, disorder, and SOC. The variation in the NFL behaviors seen the present samples may involve either UV (ultraviolet) or IR (infrared) physics. Here, UV physics refers to the appearance of localized magnetic moments near a MIT, which would become a source of strong inelastic scattering events due to their extensive entropy and thus be responsible for the NFL transport phenomena **[20]**. In contrast, IR physics means that such local moments are expected to disappear and long-wave length and low-energy fluctuations determine the NFL physics by forming either singlets or magnetic orders and reducing the huge entropy dramatically at low $T$ **[21]**. Since incompletely screened local moments play a central role in NFL behaviors, it is natural to expect an appearance of negative MR. As shown in the previous sections; however, Pv-SrIrO$_3$ films on various substrates show positive MR up to ±9 T. Although we can't ignore the role of local-moments (and related UV physics) for the variation of temperature exponents ($\varepsilon$) in resistivity, the positive MR leads us to focus on IR physics which seems to be responsible for the observed NFL behaviors.

One of the possible scenarios within IR physics is to take into account quantum Griffiths effects. In this scenario, local fluctuations between metallic and insulating islands, referred to as rare events, dominate the NFL behaviors. These rare events result from extreme inhomogeneity which may originate due to disorder. It is also important to notice that the residual resistivity is high (~mΩ·cm), implying that the concentration of disorders is not low. As a result, we conjecture that the interplay between electron correlation and disorder in the presence of strong SOC can give rise to a Griffiths-type phase between the Landau's Fermi-liquid state and the Mott-Anderson insulating phase, which allows a continuous change of the transport exponent. We would call this physics the "Mott-Anderson-Griffiths" scenario. Disorder and the evolution of NFL behaviors are possibly inter-connected as it was shown that in a correlated metal, when the disorder parameter is higher than a certain critical value, the system enters the Griffiths phase displaying NFL behaviors **[22]**. Until now, the Griffith scenario has been realized near the infinite randomness fixed point **[23]**, where extreme inhomogeneity and associated rare phenomena are responsible for NFL physics with varying critical exponents. In our case we believe that a similar situation occurs, where fluctuations between metallic and insulating islands as rare events are expected to allow the Mott-Anderson-Griffith phase, responsible for the NFL physics. Currently we are developing a theoretical model based on this Mott-Anderson-Griffith physics.

**4. Conclusions**

In conclusion, we grew high quality epitaxial Pv-SrIrO$_3$ thin films on various lattice-mismatched substrates. Intriguingly, imposing compressive strain on the film by altering the underlying lattice-mismatched substrates, results in decreases of the electronic bandwidth.

With an increase of compressive strain, the strained films show the temperature dependent resistivity $\rho \propto T^\varepsilon$ in the wide temperature range. Intriguingly, $\varepsilon$ evolves from 4/5 to 1 and to 3/2 with an increase of compressive strain. In addition, magnetoresistance remains positive irrespective of any unusual feature in resistivity. We conjecture that these results would imply a subtle interplay of correlation, disorder, and SOC. The present observations hopefully pave a way for more activities to understand this rapidly developing, yet poorly understood 5$d$ based oxide physics.

**Acknowledgements**

YHJ acknowledges the support by NRF via the Center for Topological Matter at POSTECH (Grant No. 2011-0030786). KSK was supported by NRF (Grant Nos. 2012R1A1B3000550 and 2011-0030785) and by TJ Park Science Fellowship of the POSCO TJ Park Foundation.

**References**

[1] G. R. Stewart, Non-Fermi-liquid behaviors in d- and f-electron metals, Rev. Mod. Phys. 73 (2001) 797-855. (DOI: http://dx.doi.org/10.1103/RevModPhys.73.797)

[2] E. Miranda, V. Dobrosavljević, Disorder-driven non-Fermi liquid behavior of correlated electrons, Rep. Prog. Phys. 68 (2005) 2337-2408. (DOI:10.1088/0034-4885/68/10/R02)

[3] V. Dobrosavljević, N. Trivedi, and J. M. Valles, Jr., Conductor Insulator Quantum Phase Transitions, Oxford University Press, Oxford, 2012.

[4] B. J. Kim, H. Ohsumi, T. Komesu, S. Sakai, T. Morita, H. Takagi, T. Arima, Phase-Sensitive Observation of a Spin-Orbital Mott State in $Sr_2IrO_4$, Science 323 (2009) 1329-1332. (DOI: 10.1126/science.1167106)

[5] B. J. Kim, H. Jin, S. J. Moon, J. –Y. Kim, B. –G. Park, C. S. Leem, J. Yu, T. W. Noh, C. Kim, S. –J. Oh, J. –H. Park. V. Durairaj, G. Cao, E. Rotenberg, Novel $J_{eff}$=1/2 Mott

State Induced by Relativistic Spin-Orbit Coupling in $Sr_2IrO_4$, Phys. Rev. Lett. 101 (2008) 076402. (DOI: http://dx.doi.org/10.1103/PhysRevLett.101.076402)

[6] W. W. Krempa, G. Chen, Y. –B. Kim, L. Balents, Correlated Quantum Phenomena in the Strong Spin-Orbit Regime, Condens. Matter Phys. 5 (2014) 57-82. (DOI: http://dx.doi.org/10.1146/annurev-conmatphys-020911-125138)

[7] S. J. Moon, H. Jin, K. W. Kim, W. S. Choi, Y. S. Lee, J. Yu, G. Cao, A. Sumi, H. Funakubo, C. Bernhard, T. W. Noh, Dimensionality-Controlled Insulator-Metal Transition and Correlated Metallic State in 5$d$ Transition Metal Oxides $Sr_{n+1}Ir_nO_{3n+1}$ ($n$ = 1, 2, and ∞), Phys. Rev. Lett. 101 (2008) 226402. (DOI: http://dx.doi.org/10.1103/PhysRevLett.101.226402)

[8] J. H. Gruenewald, J. Nichols, J. Terzic, G. Cao, J. W. Brill, S. S. A. Seo, Compressive strain-induced metal-insulator transition in orthorhombic $SrIrO_3$ thin films, J. Mater. Res. 29 (2014) 2491-2496. (DOI: http://dx.doi.org/10.1557/jmr.2014.288)

[9] A. Biswas, K. –S. Kim, Y. H. Jeong, Metal insulator transitions in perovskite $SrIrO_3$ thin films, J. Appl. Phys. 116 (2014) 213704. (DOI: http://dx.doi.org/10.1063/1.4903314)

[10] L. Zhang, Q. Liang, Y. Xiong, B. Zhang, L. Gao, H. Li, Y. B. Chen, J. Zhou, S. –T. Zhang, Z. –B. Gu, S. Yao, Z. Wang, Y. Lin, Y. –F. Chen, Tunable semimetallic state in compressive-strained $SrIrO_3$ films revealed by transport behavior, Phys. Rev. B. 91 (2015) 035110. (DOI: http://dx.doi.org/10.1103/PhysRevB.91.035110)

[11] Y. F. Nie, P. D. C. King, C. H. Kim, M. Uchida, H. I. Wei, B. D. Faeth, J. P. Ruf, J. P. C. Ruff, L. Xie, X. Pan, C. J. Fennie, D. G. Schlom, K. M. Shen, Interplay of Spin-Orbit Interactions, Dimensionality, and Octahedral Rotations in Semimetallic $SrIrO_3$, Phys. Rev. Lett. 114 (2015) 016401. (DOI: http://dx.doi.org/10.1103/PhysRevLett.114.016401)

[12] J. M. Longo, J. A. Kafalas, R. J. Arnott, Structure and properties of the high and low pressure forms of $SrIrO_3$, J. Solid State Chem. 3 (1971) 174-179. (DOI: http://dx.doi.org/10.1016/0022-4596(71)90022-3)

[13] G. Cao, V. Durairaj, S. Chikara, L. E. Delong, S. Parkin, P. Schlottmann, Non-Fermi-liquid behavior in nearly ferromagnetic $SrIrO_3$ single crystals, Phys. Rev. B. 76 (2007) 100402 (R). (DOI: http://dx.doi.org/10.1103/PhysRevB.76.100402)

[14] J. G. Zhao, L. X. Yang, Y. Yu, F. Y. Li, R. C. Yu, Z. Fang, L. C. Chen, C. Q. Jin,

High-pressure synthesis of orthorhombic SrIrO$_3$ perovskite and its positive magnetoresistance, J. Appl. Phys. 103 (2008) 103706.
(DOI: http://dx.doi.org/10.1063/1.2908879)

[15] P. E. R. Blanchard, E. Reynolds, B. J. Kennedy, J. A. Kimpton, M. Avdeev, A. A. Belik, Anomalous thermal expansion in orthorhombic perovskite SrIrO$_3$: Interplay between spin-orbit coupling and the crystal lattice, Phys. Rev. B. 89 (2014) 214106.
(DOI: http://dx.doi.org/10.1103/PhysRevB.89.214106)

[16] P. A. Lee, T. V. Ramakrishnan, Disordered electronic systems, Rev. Mod. Phys. 57 (1985) 287-337. (DOI: http://dx.doi.org/10.1103/RevModPhys.57.287)

[17] H. Yamada, S. Takada, Negative Magnetoresistance of Ferromagnetic Metals due to Spin Fluctuations, Prog. Theor. Phys. 48 (1972) 1828-1848.
(DOI: http://dx.doi.org/10.1143/PTP.48.1828)

[18] W. A. Harrison, Electronic structure and the physical Properties of Solids: The Physics of the Chemical Bonds, New York, 1989.

[19] A. Kawabata, Theory of negative magnetoresistance in three-dimensional systems, Solid State Commun. 34 (1980) 431-432. (DOI:10.1016/0038-1098(80)90644-4)

[20] A. Georges, G. Kotliar, W. Krauth, M. J. Rozenberg, Dynamical mean-field theory of strongly correlated fermion systems and the limit of infinite dimensions, Rev. Mod. Phys. 68 (1996) 13-125. (DOI: http://dx.doi.org/10.1103/RevModPhys.68.13)

[21] H. v. Löhneysen, A. Rosch, M. Vojta, P. Wölfle, Fermi-liquid instabilities at magnetic quantum phase transitions, Rev. Mod. Phys. 79 (2007) 1015-1075.
(DOI: http://dx.doi.org/10.1103/RevModPhys.79.1015)

[22] V. Dobrosavljević, G. Kotliar, Mean Field Theory of the Mott-Anderson Transition, Phys. Rev. Lett. 78 (1997) 3943-3946.
(DOI: http://dx.doi.org/10.1103/PhysRevLett.78.3943)

[23] T. Vojta, Phases and phase transitions in disordered quantum systems, arXiv:1301.7746.

**Table 1**

Temperature dependent resistivity variation of epitaxial Pv-SrIrO$_3$ thin films grown on various substrates. '+' sign corresponds to the amount of compressive strain.

| Substrate | Strain | $\rho \propto T^{\varepsilon}$ | Low $T$ phenomena |
|---|---|---|---|
| GdScO$_3$ (110) | 0.00% | $\varepsilon = 4/5$ | Fermi-liquid at $T \leq 10$ K |
| DyScO$_3$ (110) | +0.50% | $\varepsilon = 1$ | Resistivity upturn at $T \leq 20$ K |
| SrTiO$_3$ (001) | +1.54% | $\varepsilon = 3/2$ | Resistivity upturn at $T \leq 50$ K |

**Figure Captions**

**Fig. 1.** (Color Online) (a) *Pseudo-cubic* (or cubic) lattice constants of perovskite $SrIrO_3$ and substrates $GdScO_3$ (110), $DyScO_3$ (110), and $SrTiO_3$ (001). Corresponding amount of compressive strain is also shown. (b) X-ray $\theta$-$2\theta$ scan of epitaxial $SrIrO_3$ thin films of 35 nm thickness grown on three different substrates. Bragg peaks and thickness fringes are seen. Only low angle *pseudo-cubic* $(001)_{pc}$ peaks are shown for clarity.

**Fig. 2.** (Color Online) Temperature dependent electrical resistivity of epitaxial $SrIrO_3$ thin films grown on three different substrates. With increasing compressive strain, resistivity at $T = 300$ K increases. Best lattice-matched film on $GdScO_3$ shows a fully metallic behavior down to lowest temperature. For the strained films, low temperature upturns in resistivity ($T = 20$ K for films on $DyScO_3$ and $T = 50$ K for films on $SrTiO_3$) are indicated by arrow.

**Fig. 3.** (Color Online) Fitting the temperature dependent resistivity to a power law $\rho \propto T^{\varepsilon}$. Resistivity in the metallic region follows the power law: $\rho \propto T^{4/5}$ for film on $GdScO_3$; $\rho \propto T$ for film on $DyScO_3$; and $\rho \propto T^{3/2}$ for film on $SrTiO_3$.

**Fig. 4.** (Color Online) Low temperature upturn regions in resistivity ($T \leq 20$ K for films on $DyScO_3$ and $T \leq 50$ K for films on $SrTiO_3$) are fitted to weak localization theory, $\sigma \propto \ln T$.

**Fig. 5.** (Color Online) Magnetoresistance (MR) of $SrIrO_3$ thin films grown on $GdScO_3$, $DyScO_3$, and $SrTiO_3$ substrates. Transverse MR (field perpendicular to the plane) is positive and quadratic at $T = 5$ K for all three films up to $\pm 9$ T.

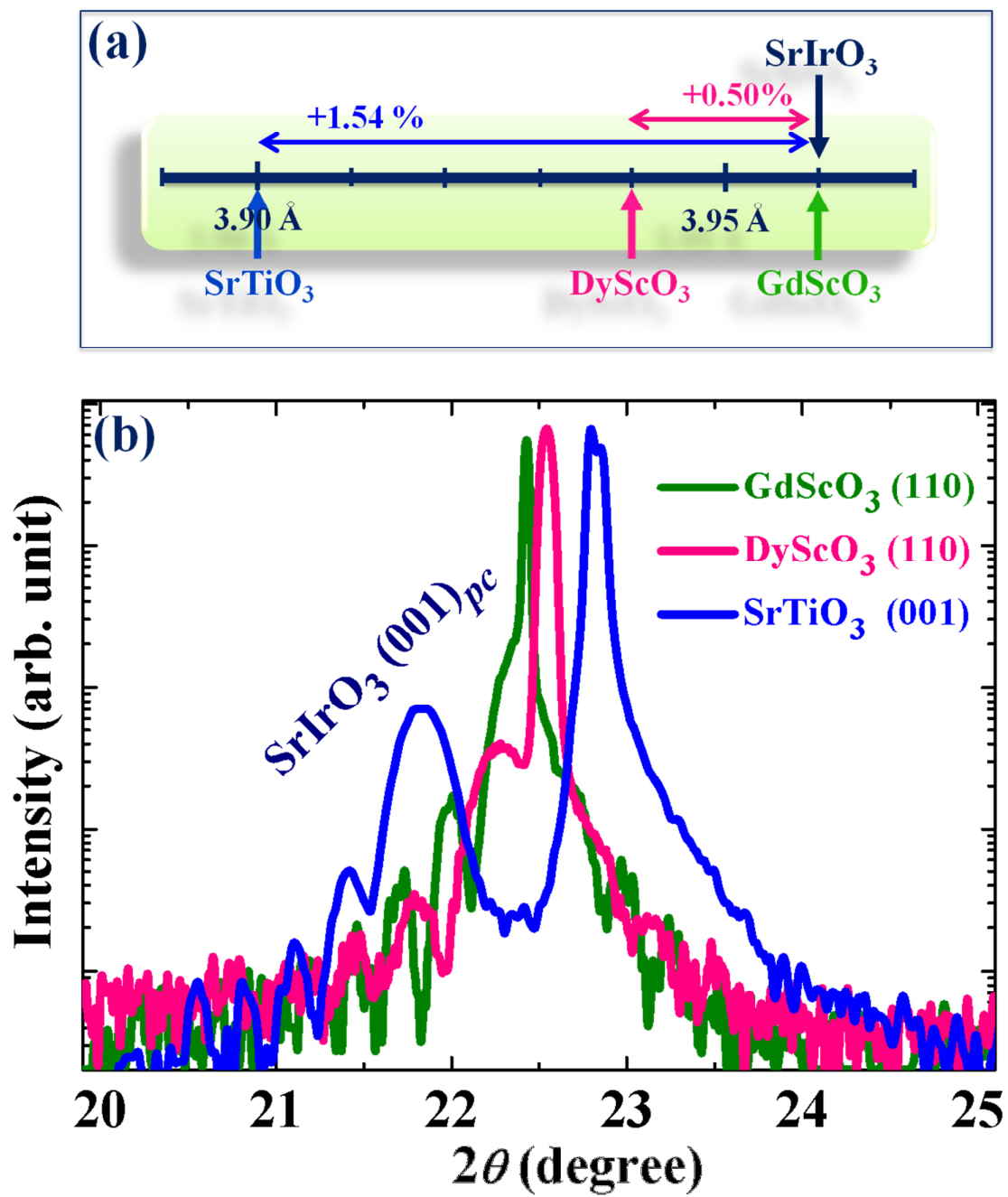

(Figure 1)

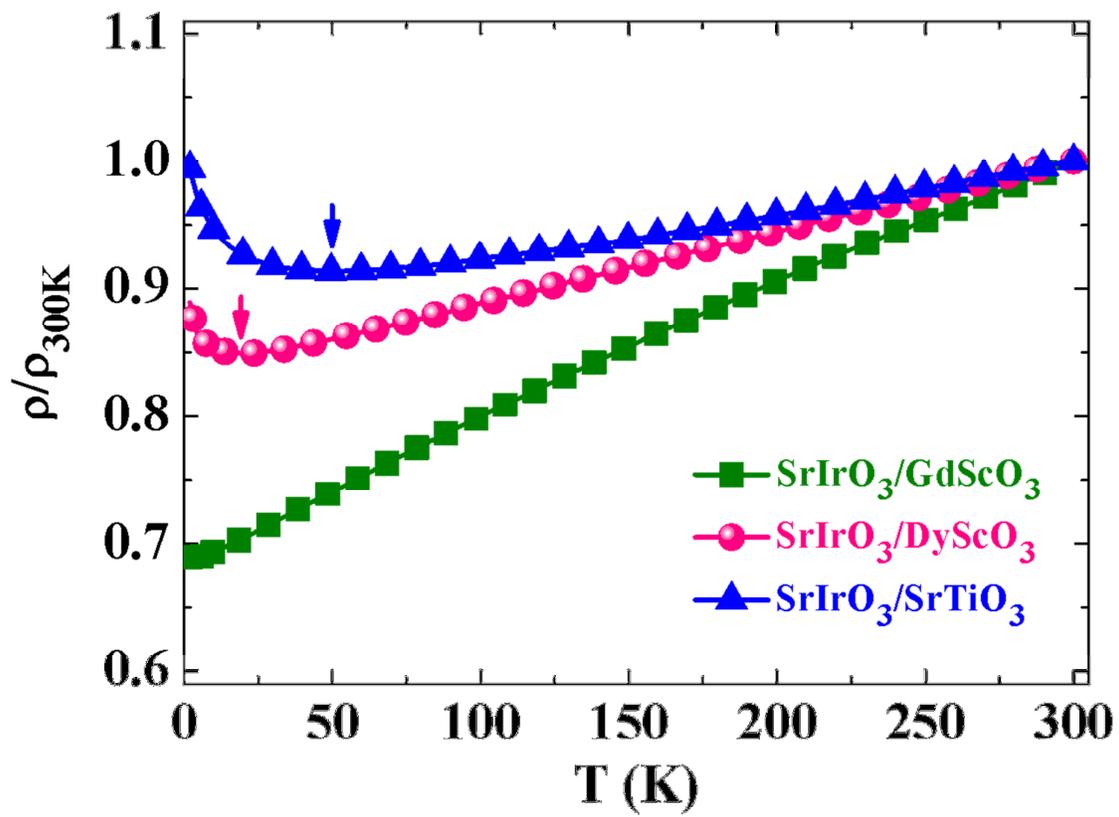

(Figure 2)

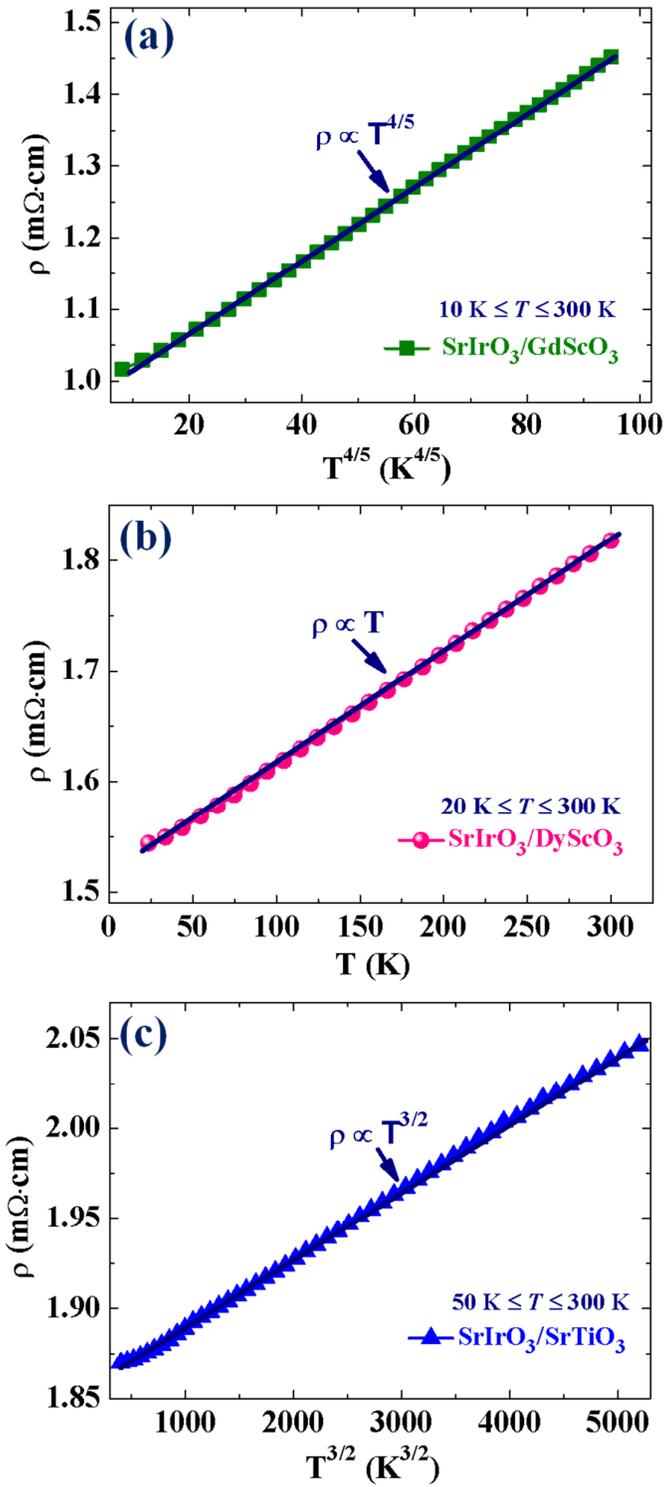

(Figure 3)

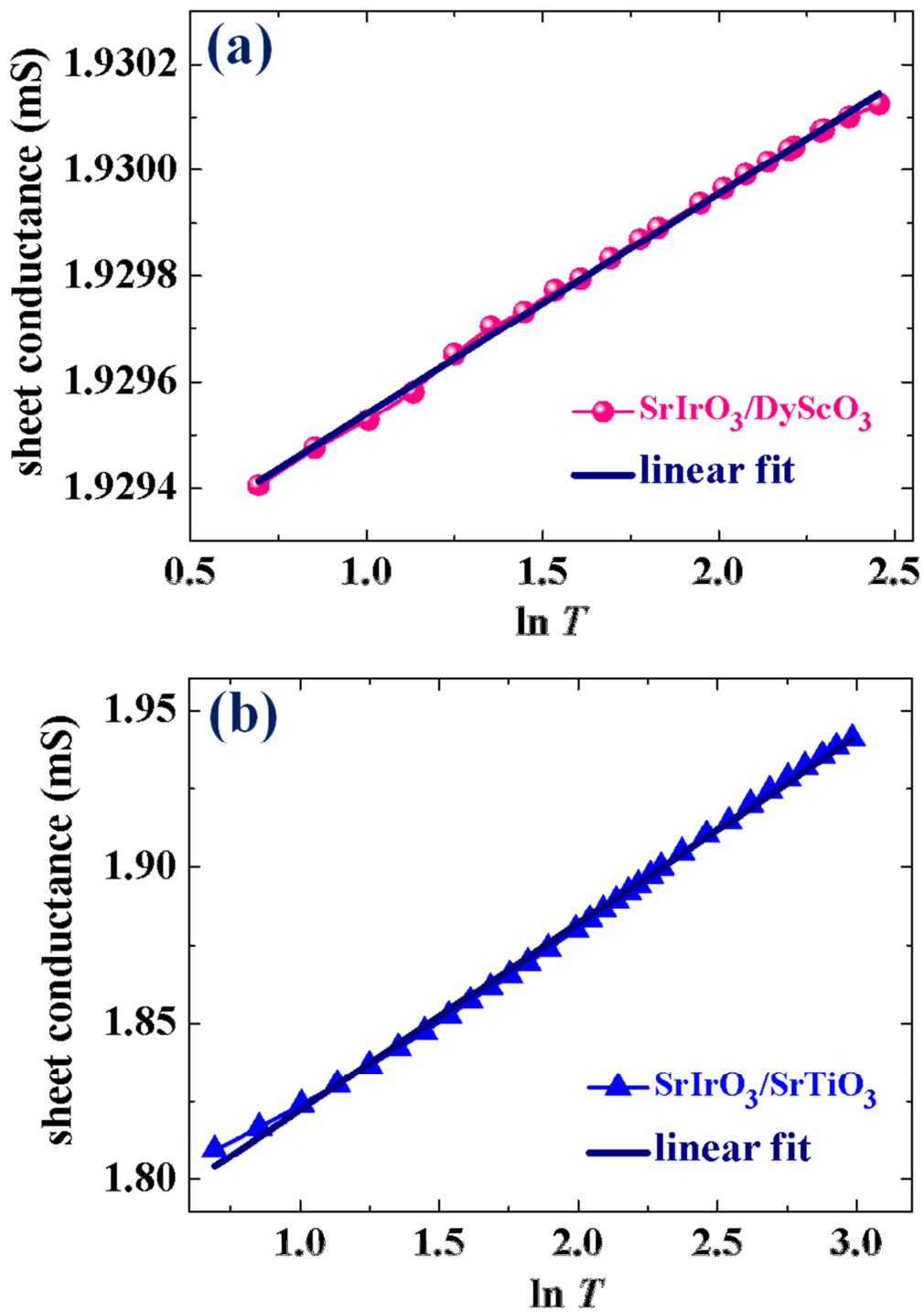

(Figure 4)

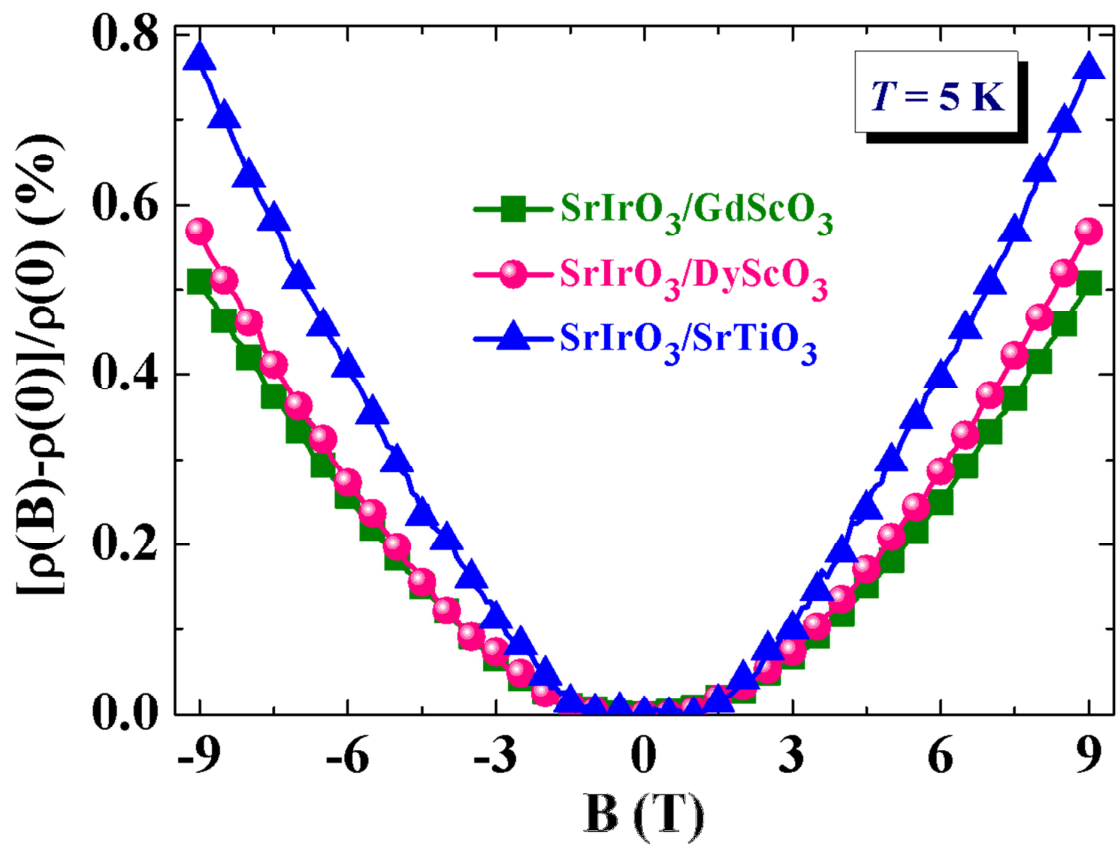

**(Figure 5)**